\documentclass[apj]{emulateapj}

\usepackage{natbib}
\newcommand{\msini}{\ensuremath{m \sin{i}}}
\newcommand{\feh}{\ensuremath{[\mbox{Fe}/\mbox{H}]}}
\newcommand{\teff}{\ensuremath{T_{\mbox{\scriptsize eff}}}}

\newcommand{\Nstars}{363}
\newcommand{\Nplanets}{427}

\begin{document} 

\title{The Exoplanet Orbit Database}

\author{J. T. Wright\altaffilmark{1,2}, O. Fakhouri\altaffilmark{3,4}, G. W. Marcy\altaffilmark{4,5}, E. Han\altaffilmark{1,2}, Y. Feng\altaffilmark{1,2}, John Asher Johnson\altaffilmark{6}, A. W. Howard\altaffilmark{4,7}, D. A. Fischer\altaffilmark{8}, J. A. Valenti\altaffilmark{9}, J. Anderson\altaffilmark{9}, N. Piskunov\altaffilmark{10}}  %, H. Issacson\altaffilmark{6}}

\altaffiltext{1}{Center for Exoplanets and Habitable Worlds, 525 Davey Lab, The Pennsylvania State University, University Park, PA 16803}
\altaffiltext{2}{Department of Astronomy \& Astrophysics, 525 Davey Lab, The Pennsylvania State University, University Park, PA 16803}
\altaffiltext{3}{Pivotal Labs, 731 Market Street, Third Floor, San Francisco, CA 94103}
\altaffiltext{4}{Department of Astronomy, University of California, Berkeley, CA 94720-3411, USA} 
\altaffiltext{5}{Center for Integrative and Planetary Science, University of California, Berkeley, CA 94720}
\altaffiltext{6}{Department of Astrophysics, California Institute of Technology, MC 249-17, Pasadena, CA 91125, USA}
\altaffiltext{7}{Space Sciences Laboratory, University of California, Berkeley, CA 94720-7450 USA}
\altaffiltext{8}{Department of Astronomy, Yale University, New Haven, CT 06511, USA}
\altaffiltext{9}{Space Telescope Science Institute, 3700 San Martin Dr., Baltimore, MD 21218, USA}
\altaffiltext{10}{Department of Astronomy and Space Physics, Uppsala University, 
                        Box 515, 751 20 Uppsala, Sweden}

 \begin{abstract}

We present a database of well determined orbital parameters of exoplanets, and their host stars' properties.  This database comprises spectroscopic orbital elements measured for \Nplanets\ planets orbiting \Nstars\ stars from radial velocity and transit measurements as reported in the literature.  We have also compiled fundamental transit parameters, stellar parameters, and the method used for the planets discovery.  This Exoplanet Orbit Database includes all planets with robust, well measured orbital parameters reported in peer-reviewed articles.  The Database is available in a searchable, filterable, and sortable form on the Web at http://exoplanets.org through the Exoplanets Data Explorer Table, and the data can be plotted and explored through the Exoplanet Data Explorer Plotter.  We use the Data Explorer to generate publication-ready plots giving three examples of the signatures of exoplanet migration and dynamical evolution:  We illustrate the character of the apparent correlation between mass and period in exoplanet orbits, the different selection biases between radial velocity and transit surveys, and that the multiplanet systems show a distinct semi-major axis distribution from apparently singleton systems. 

 \end{abstract}

\section{Introduction}  
Since the first discovery of exoplanets orbiting normal stars \citep{Latham89,Mayor_queloz} the number of known exoplanets has grown rapidly, predominantly through the precise radial velocity (RV) method.  Recently, exoplanet discoveries via transit have begun to keep pace, and the {\it Kepler} mission to detect transiting planets promises to surpass RV methods, and other methods such as microlensing and direct imaging have made promising progress.  Careful tracking of the many dozens of discoveries per year has been carried out by a few groups, most notably the {\it Extrasolar Planet Encyclopedia} maintained at {\tt http://exoplanet.eu} by Jean Schneider, and more recently the  NASA/NExScI/IPAC Stellar and Exoplanet Database (NStED) at {\tt http://nsted.ipac.caltech.edu}.  

The first peer-reviewed list of exoplanets with robust orbits appearing in the peer-reviewed literature was appeared in \citet{Butler02}.  \citet{Fischer05c} compiled a comprehensive list of uniformly calculated orbital parameters and stellar properties for planets orbiting stars monitored by the California \& Carnegie and Anglo-Australian Planet Searches. 

In {\it Catalog of Nearby Exoplanets}, \citep{Butler06} presented orbital and stellar parameters for the 172 exoplanets with well determined orbits around normal stars known within 200 pc.  At that time, only a handful of planets had been discovered through the transit method, and the distance threshold served to distinguish planets orbiting the brightest and most easily studied stars from more distant planets around faint stars with ill-determined orbits, such as the planets discovered by microlensing.

We have maintained and updated the {\it Catalog}, and have expanded it to include additional information, including transit parameters and asymmetric uncertainties.  We have made this Exoplanet Orbit Database available online and developed the Exoplanet Data Explorer to allow users to easily explore and display its contents.  This article serves to document the methodology of the {\it EOD} and subject it to peer review.  We anticipate many future upgrades to the {\it EOD}, including the addition of fields not currently supported and more thorough documentation of references.

\section{Scope and Purpose}

For the Exoplanet Orbit Database, we have dropped the 200 pc limit from the old {\it Catalog}, and now include all robustly detected planets appearing in the peer-reviewed literature with well determined orbital parameters.  We have retained the generous upper mass limit of 24 Jupiter masses in our definition of a ``planet'', for the same reasons as in the {\it Catalog}: at the moment, any mass limit is arbitrary and will serve little practical function both because of the $\sin i$ ambiguity in radial velocity masses and because of the lack of physical motivation.\footnote{The 13 Jupiter-mass limit by the IAU Working Group is physically unmotivated for planets with rocky cores, and observationally unenforceable due to the sin {\it i} ambiguity.   A useful theoretical and rhetorical distinction is to segregate brown dwarfs from planets by their formation mechanism, but such a distinction is of little utility observationally.}  We therefore err on the side of inclusiveness by admitting the long high-mass tail of the exoplanet population at the risk of having a few {\it bona fide} brown dwarfs in the sample.  

The scope of this Exoplanet Orbit Database (EOD) is to provide the highest quality orbital parameters for exoplanets orbiting ``normal'' stars.  We are not attempting to provide an encyclopedic presentation of every claimed detection of an exoplanet.\footnote{This service is admirably provided by the Extrasolar Planet Encyclopedia.  Since this task becomes more complex as new planet detection methods explore new dimensionalities of exoplanet observation, we restrict ourselves to orbital parameters determined spectroscopically, supplemented in the cases of transits with photometry.}  At present, we include giant and subgiant stars because exoplanet detection methods and measurement uncertainties for these stars are similar to main-sequence stars. In the future, we may include other evolutionary states such as hot subdwarfs, white dwarfs, post-CE binaries, or pulsars.  We plan to include astrometrically discovered planets when they appear in the literature with robust orbital elements.  
 
Our definition of ``robust'' is not strictly quantitative.  We require that the period be certain to at least 15\% (usually corresponding to seeing at least one or two complete orbits), but otherwise we have applied our judgement regarding whether both the detection and the orbit are sufficiently secure to warrant inclusion in the Database.  We attempt to be conservative in these evaluations.  Our standards for the quality of a radial velocity curve might, for instance, be relaxed if a given planet transits, or tightened if phase coverage is especially poor.  In any case we strive to avoid including dubious orbits or detections that we may need to revise at a later date.  We stress that this judgement is not necessarily a judgement on the quality of other groups' work generally or the existence of a particular planet --- indeed we have not included some very real planets published in our own manuscripts because their orbital parameters are not sufficiently well determined to meet the Database's standards. 

We also collect basic information regarding the quality of the orbital fit, including the number of velocity measurements made, the r.m.s. scatter about the fit, and the resulting $\chi^2$.  Finally, we collect substantial auxiliary information regarding the host star, including its best measured parallax, mass, and activity levels.  We provide references for nearly all quantities, and our website provides easy links to these refereed sources.

Thus, the EOD provides added value to other compendia of exoplanet properties in that:
\begin{itemize} 
\item it provides a ``quality cut'', containing only robust orbital parameters for clearly detected planets appearing in the peer-reviewed literature;
\item it distinguishes derived quantities, such as \msini\, from measured quantities, such as period, eccentricity, and RV semiamplitude (the last of which, for instance, is not stored in other compendia).  This allows derived quantities to be recalculated when, for instance, better stellar masses become available.  
\item it provides a suite of stellar and orbital fit parameters, such as the number of radial velocity observations in the fit, the quality of the published fit, and the mass, projected rotational velocity, and chromospheric activity level of the host star;
\item it  links to the underlying radial velocity and photometric data that generated the orbital fit;
\item it is available on a website that provides a powerful and visually elegant data exploration and visualization tool.
\end{itemize}

We stress that the heterogenous detection thersholds within and amongst the many exoplanet search programs responsible for the detection and characterization of the known exoplanets make a sensitive analysis of the global properties of the known exoplanet treacherous.  An obvious example is the very different properties of the host stars and orbits of planets discovered by transit versus those discovered by RVs.  While this particular factor can be crudely addressed through use of the DISCMETH field in the EOD, other factors are less obvious and more difficult to control.  A more subtle example is that the cadence and radial velocity precision achieved on particular targets by the many telescopes, groups, and techniques varies as a function of stellar spectral type, as a function of magnitude, and in less predictable ways.  {\em Thus, careful consideration of the many and often ill-defined selection effects in planet search programs is crucial when interpreting these data statistically to find astrophysically meaningful correlations or effects.}

\section{Content}

Our methodology largely follows that of \citet{Butler06}.  We summarize the important points and differences from that work below.

\subsection{Data}
The data in the EOD are stored in flat text files, one per planet.  Below, we describe each of the fields and how we determine its value.  The names of the fields as used in the Database are specified in all CAPS in the text below, and are summarized in Table~\ref{table}.

We record the published fundamental observables of SB1's:  period ($P$, stored as PER), semiamplitude ($K$), eccentricity ($e$, stored as ECC), and the time and argument of periastron ($T_0, \omega$, stored as T0 and OM), and their uncertainties.  In a few cases of multiplanet systems for which orbital parameters are not constant over the span of the observations, we report the osculating elements at the epoch given in the source.  We also record the presence of a linear trend (TREND) and its magnitude (DVDT), where relevant, and whether the eccentricity was frozen in the orbital fit (FREEZE\_ECC).  In the case of circular orbits, we choose $T_0=T_{\mbox{t}}$ and $\omega=90^\circ$, where $T_{\mbox{t}}$ is the epoch of transit center.

We have opted to use these ``classical'' SB1 orbital parameters rather than using mean longitude at epoch because they are more frequently reported in the literature and the the latter is trivially computed from the former.  In those cases (especially for multiplanet systems or transiting systems) where the phase of a planet is reported as $M$, or epoch of transit center, or in some similar way, we have converted the quantities to $\omega$ and $T_0$ for consistency.  We recognize that for circular orbits the uncertainty in mean longitude is better behaved than those in $T_0$ and $\omega$, and we note that the uncertainty in mean longitude can be estimated from the period uncertainty and the span of the observations.  We plan to incorporate mean longitude at epoch, transit time predictions, and robust uncertainties for these quantities in the future, but in the meantime any application requiring more precision should calculate the quantity explicitly from the radial velocities or from the source manuscript.

We have attempted to make the stellar mass measurements as uniform as possible, with many masses coming from \citet{Takeda07} instead of the planet discovery articles.  From the five orbital parameters and these masses, we calculate the minimum mass \msini\ (MSINI) and the orbital semimajor axis $a$ (A) for every planet following the methodology of \citet{Wright09b} and \citet{Butler06}. Note that because we often use stellar masses that differ from the discovery manuscript values, the minimum masses and $a$ values may differ from their discovery values.  In articles where the minimum mass of planets are given but not $K$ (for instance, in multiplanet systems where only a dynamical fit is given), we have computed $K$ from the $M_*$, $P$, $e$, and $M\sin i$ values in the Database for consistency.

We report stellar parallaxes (PAR) and coordinates using the rereducion of the Hipparcos dataset by \citet{Hipparcos2}, where available, and from discovery articles otherwise.\footnote{In a few cases we have had to estimate distances directly from stellar parameters;  in these cases we have attempted to be conservative in our error estimates.}    Coordinates are stored in the RA and DEC fields as decimal quantities, and in RA\_STRING and DEC\_STRING as sexigesimal strings.   The V and BMV fields contain the $V$ magnitude and $B-V$ color, usually from the Hipparcos catalog \citet{PerrymanESA}, and $JHK_{\mbox{S}}$ photometry is from 2MASS \citep{2MASS} (contained in the fields J, H, and KS, the latter being distinguished from the semiamplitude K).  For stars not appearing in those catalogs, the values come from the discovery articles.   Chromospheric activity measurements are from the discovery articles, or from the values listed in \citet{Butler06}, and are stored as Mount Wilson S values (SHK) and $\log R^\prime_{\mbox{HK}}$ (RHK).  

Where the literature is not consistent, we use proper names, Bayer designations, or Flamsteed numbers to identify a star in the STAR field, where available, because we find those to be more mnemonic than catalog numbers. We then give priority to GJ numbers before HD numbers, and HD numbers before Hipparcos designations.  In cases where the literature violates this scheme or is inconsistent, we give an alternative name in the OTHERNAME field.  We include fields in the Database for HD numbers, HR numbers, Gliese numbers (GL), Hippacos number (HIPP), SAO number.  For Bayer designations we spell out the Greek letter component, and in all cases we use three-letter constellation abbreviations.  We provide a component name (COMP, i.e. ``b'', ``c'', ``d'', etc.), and combine the STAR and COMP fields to generate the NAME of the planet.

As in the case for stellar masses, we attempt to record as consistent a set of metalicities (FE), effective temperatures (TEFF), gravities (LOGG), and projected rotational speeds (VSINI) as possible, relying heavily on the SPOCS catalogs \citep[e.g.]{SPOCS} and studies by the Geneva group \citep[e.g.][]{Santos03_metal}.  In most other cases these values come from the discovery articles, and for the host stars of transiting planets, we prefer the $\log g$ value determined with the transit light curve to a value determined from spectroscopy alone.  We have collected spectral types from discovery articles and SIMBAD and store the values in SPTYPE, although this field is difficult to maintain or check in a consistent way.  

Stars identified as binaries in the literature have the BINARY flag set to 1.  For multiplanet systems we set the MULT flag to 1 and record the number of planets in the NCOMP field.

For planets that transit (for which the TRANSIT field is set to 1), we incorporate data on the period, epoch of transit center ($T_{\mbox{t}}$, stored as TT), impact parameter ($b$, as B), the square of the planet-star radius ratio as a DEPTH $(R_{\mbox{p}}/R_*)^2$, the time of transit from first to fourth contact ($T_{14}$ as T14), inclination ($i$, as I), orbital distance to stellar radius ratio $a/R_*$ (as AR), and planetary radius ($r$, as R).  Unlike the SB1 orbital parameters, this set is overdetermined, and we do not calculate any of these transit parameters from the others (except in cases where a parameter is not reported, and in no case do we attempt to calculate values directly from light curves).  We also record  the bulk density of the planet ($\rho$, as DENSITY).  Where these quantities are not published for a transiting planet, we have calculated them from the other parameters for completeness.  Since \msini\ is derived including the stellar mass, which may come from a different source than the reference providing the transit parameters, this may cause minor inconsistencies between the EOD and rigorously calculated values from the discovery data.  We also record the projected spin-orbit misalignment $\lambda$ (as LAMBDA, sometimes reported in the literature in terms of $\beta = -\lambda$), as measured by the Rossiter-McLaughlin effect.  We calculate planetary surface gravity ($\log g$ as GRAVITY) from the recorded transit parameters and A, using the formalism of \citet{Southworth07}, and UGRAVITY through a formal propagation of errors assuming no covariances.

In a small number of cases, it is obvious based on the data presented in planet discovery articles that the orbital parameters are misreported.  In cases where it appears to be a simple typographical error, we have simply corrected the value; in most cases the problem is a misreported offset to the Julian Date of the time of periastron passage.

We also record the method of discovery of a planetary system, DISCMETH.  This field at present can take two values: ``RV'' or ``Transit''.   So, for instance, HD 209458$b$ (which was discovered in the course of RV surveys and later found to transit) has TRANSIT$=1$ but DISCMETH=``RV'', while HAT-P-13$c$ (which is not known to transit and was discovered in the course of radial velocity follow-up for the transiting planet HAT-P-13$b$) has TRANSIT$=0$ and DISCMETH=``Transit''.  This allows for some crude corrections to the very different selection effects of RV and transit surveys in analyses of global exoplanet properties \citep[e.g.][]{Gaudi05a,Gaudi05b}.

\subsection{Uncertainties}
Where possible, we have recorded the uncertainties  from the literature, where they are computed in a nonuniform way.  Where available or trivially computed, we record the quality of the orbital fit, including the $\chi^2_\nu$ (CHI2) and r.m.s.\ residuals of the fit (RMS), and the number of RV observations used in the fit (NOBS).  

All uncertainties are stored in fields beginning with a U and followed by the field name.  Thus, the period uncertainty is specified in the field UPER.  For those fields where asymmetric uncertainties are commonly found in the literature, we record the uncertainty field as half of the span between the upper and lower limits of the uncertainty interval and we store the asymmetry in an additional field, with ends in D, as the value of the upper uncertainty.  For instance, the quantity $e=0.5^{+0.1}_{-0.2}$ would be stored as three fields:  ECC=0.5, UECC=0.15, and UECCD=0.1.  For symmetric uncertainties in the eccentricity, UECCD is undefined (or, equivalently, equal to UECC).

In many cases we have computed quantities from other literature values (e.g.\ \msini, GRAVITY, or $T_0$ for planets where only $T_{\mbox{t}}$ is given), and have had to make estimates of the uncertainties in these quantities.  In all cases we attempt to be conservative in our estimates to avoid the false precision that can come from a lack of knowledge of the covariance between quantities when propagating errors.  We have, for instance, conservatively assumed a minimum uncertainty of 5\% on all stellar masses, regardless of the formal uncertainties in the literature, to account for likely systematic effects \citep[but this may be too conservative, see][]{Torres10}.  In particular, the actual uncertainties in the surface gravities or semimajor axes of transiting planets may be lower than we report.

\subsection{References}
We provide references (REFs) for most numbers in the Database.  We do this as a simple text sting of the form ``First\_Author Year'' referring to the article from which we collected the quantity.  For instance, a reference to this manuscript would be rendered as the string ``Wright 2011''.   We also provide a URL to the Astrophysics Data System webpage of that article.  In the case of recently announced planets where an ADS page is not available, we provide a link to the relevant peer-reviewed preprint at the arXiv\footnote{http://xxx.lanl.gov}.  We provide references and URLs for the spectroscopic orbital elements in the fields ORBREF and ORBURL, respectively.  MASSREF and MASSURL contain the reference for the stellar mass, and DISTREF and DISTURL refer to the distance to the star.  SPECREF and SPECURL provide a reference for the stellar parameters such as \feh\ and \teff, and TRANSITREF and TRANSITURL refer to the  article from which we have collected transit parameters.  BINARYREF and BINARYURL contain an example of a reference to the multiplicity of a star for all stars with BINARY$=1$.  In cases where we have combined data from multiple sources, we separate the references and URLS with semicolons.  In the future we will provide references to all of the quantities in the Database, including magnitudes and coordinates.

We also provide a reference to the ``first'' peer-reviewed appearance of each planet in the literature  (FIRSTREF and FIRSTURL) for historical use, along with the year of this reference's publication (DATE).  Care should be taken with this field since many planets were first announced as tentative detections in the literature, in conference proceedings, or in a few cases by press release.    As a result this field should not be used to determine ``credit'' or priority for a planet's discovery, since in a few cases the first peer-reviewed article on a planet was not written by its discoverers, and in any event many planets effectively have co-discoverers.\footnote{A thorough, though somewhat out-of-date, compendium of planet discovery claims is available on the Web at {\tt http://obswww.unige.ch/$\sim$naef/who\_discovered\_that\_planet.html}.}

We provide the names used by the Extrasolar Planet Encyclopedia (JSNAME), NStED (NSTEDID), SIMBAD (SIMBADNAME), and the Exoplanet Transit Database (ETDNAME) for cross-referencing purposes.  

\section{Website}

\subsection{exoplanets.org}
A snapshot of the complete Database is available in the electronic version of this article, and at {\tt http://exoplanets.org} as a comma separated value file.  The website will be regularly maintained to include new planets as they are published in the literature.  Reports of errors and omissions are welcome by email at the addresses listed on the website.  We anticipate that the incorporation of new planets may have a modest delay from the date of publication to allow for confirmation that a planet is peer reviewed, careful consideration of the robustness of the orbit, and in some cases for followup or confirming observations.

When using the Database or its products in publication, it is appropriate to cite this manuscript and to include an acknowledgement similar to ``This research has made use of the Exoplanet Orbit Database and the Exoplanet Data Explorer at exoplanets.org,'' as appropriate.

\subsection{The Data Explorer}

The EOD can be explored and displayed using the Exoplanet Data Explorer Table and Plotter.  

\begin{figure*}
\plottwo{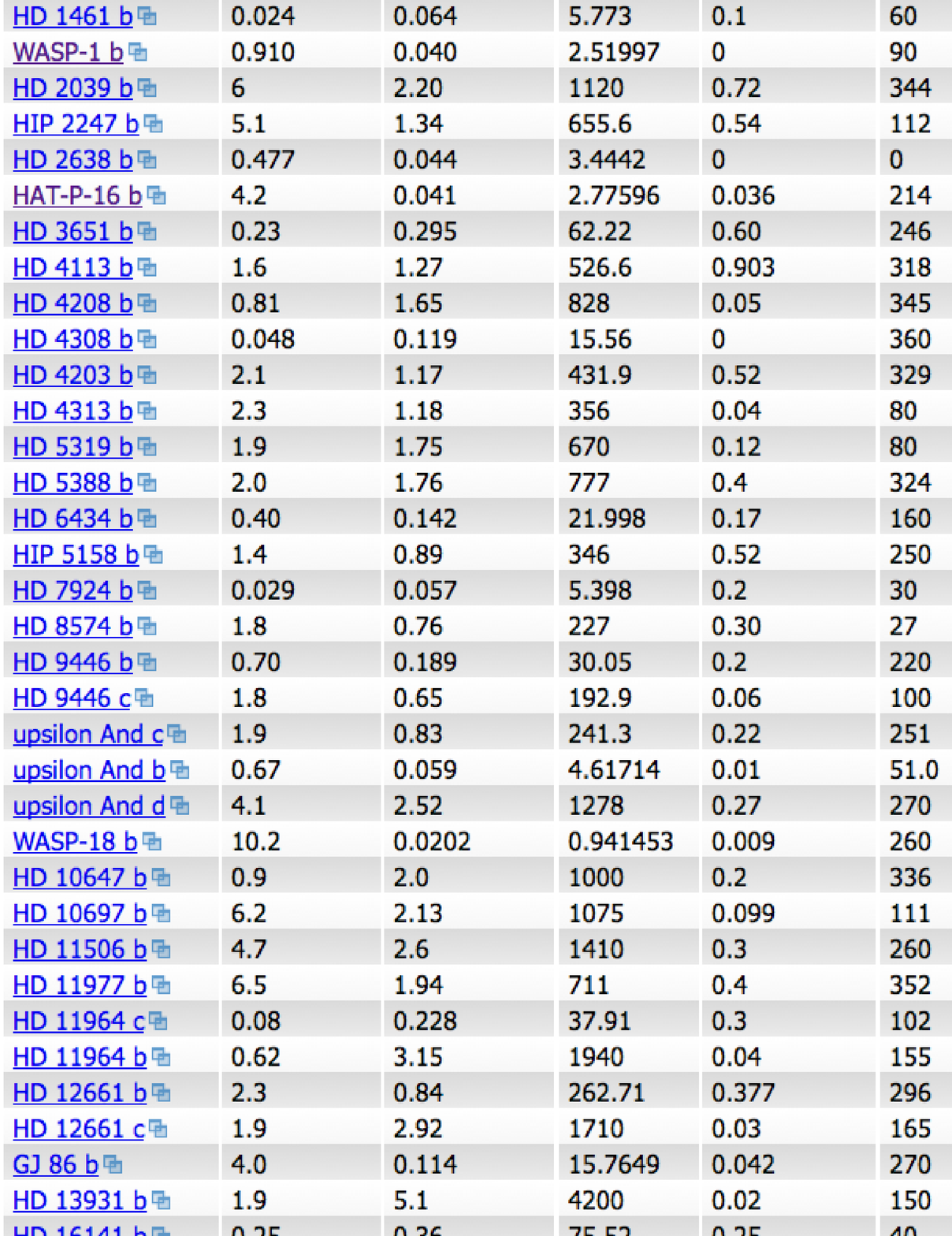}{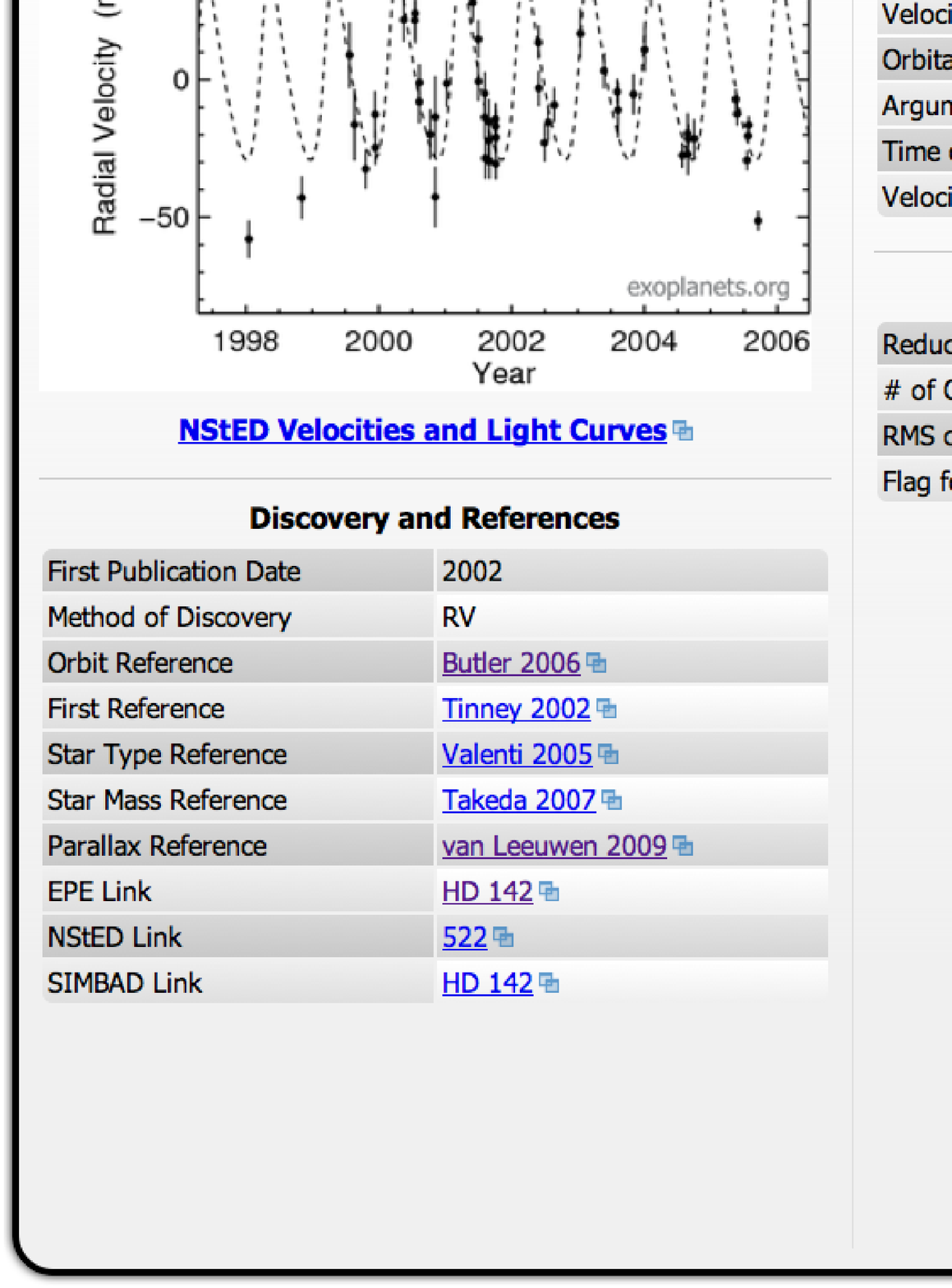}
\caption{An example of the table interface (left) and a ``one-up'' page (right).\label{oneoff}}
\end{figure*}

The Table Explorer allows for the user to dynamically create a sorted table of planets and selected properties, including a choice of units and parameter uncertainties.  Once a table has been generated, it may be exported as a custom text file.  References are linked to their corresponding URLS, we provide columns for links to SIMBAD, NStED, and the ETD, and planets are linked to ``one-up'' planet pages which contain all fields and values for a given set of planets.    Both pages as illustrated in Figure~\ref{oneoff}.

These ``one-up'' pages include a link to the publicly available velocities of each star, stored at NStED, and a plot showing these published velocities as a function of time or phase (as appropriate) along with a velocity curve generated from the listed orbital solution.  Note that we have not attempted to fit the velocities and generate our own solution;  we solve only for the velocity offset $\gamma$ and simply overplot the solution and data.  This serves as a check on accuracy of our transcription of orbital elements.  

The Plotter Explorer allows for the quantitative fields to be plotted as scatterplots or histograms, including asymmetric error bars, logarithmic axes, annotated axes, custom axis ranges, plot symbol sizes and styles, and line widths.  It also allows for additional quantities to be displayed as color-coding of plotted symbols or symbol sizes, and for multiple charts to appear overplotted in different colors (especially useful for histograms).  Plot axes and error bars can be specified with arbitrarily complex formulae using any field in the EOD (see Fig.~\ref{transit} for a simple example). 

These tables and plots can be performed on any subset of the Database through the use of filters.  These filters can be arbitrarily complex, including restrictions on arithmetically combined parameters (for instance, one could search for all RV-discovered planets whose periods are known to better than 5\% through the filter {\tt UPER/PER < 0.05 AND DISCMETH =`RV'}) and with a variety of units (units are accessed with square brackets:  {\tt MSINI[mjupiter]} or {\tt MSINI[g]} for grams).  Filters and plot settings can be saved for future use, as described below, so that plots can be regenerated at a later time with the latest version of the EOD without rebuilding the plot ``by hand''.

Plots can be exported in several formats, including PNG, SVD, and PDF, and in an arbitrary aspect ratio.  We also provide suggested output settings for presentation-quality plots (e.g. for PowerPoint) and for publication.  Users can then further annotate plots using their own presentation software, or download the data used to generated the plot (through the filter and export features of the EDE Table) and use their favorite plotting software to make a custom plot.

\subsection{Implementation of the Data Explorer}
The Exoplanet Data Explorer is a web application that aims to make data analysis in the web browser possible, practical, and accessible. This is accomplished by transferring as much of the data processing load as possible from the server onto the user's browser, and by leveraging the latest browser standards (commonly referred to as the HTML 5 standards) to give users a rich, low-latency, environment to manipulate the EOD.

The server code is implemented using the Python programming language and exists solely to provide the front-end client (the browser) access to the underlying data stored on the server in a {\tt sqlite} Database. The client code is a mix of HTML for document layout, CSS for document styling, and JavaScript for program logic. JavaScript, not to be confused with Java, is a programming language introduced by the Netscape Communications Corporation in 1995 to facilitate the production of dynamic web pages; despite many misconceptions, JavaScript is a full-fledged, mature, object-oriented language capable of building complex applications.

JavaScript is used to construct the Data Explorer's rich interactive user interface. Table columns are draggable and sortable, units and errors can be toggled via drop-down menus, the set of available planet properties can be quickly searched to pinpoint the desired property in real time -- all of this functionality is provided by JavaScript. In fact, the interface components themselves are implemented using a custom JavaScript driven GUI framework to allow for a consistent, customizable, look and feel across browsers. We use a small number of external libraries; of these the most important is the open source jQuery library\footnote{{\tt http://jquery.com}} which provides a thoughtful and consistent cross-browser Application Programming Interface for manipulating HTML elements.

We also use JavaScript to write a custom language parser based on Douglas Crockford's implementation \citep{Crockford07} of a top down operator precedence parsing algorithm first described by \citet{Pratt73}. This parser allows the user to construct and apply arbitrarily complex cuts on the EOD dataset using a simple, but powerful, query language. Since these filters are parsed in the browser they can be modified in real time without the delay commonly associated with queries that must make the round trip between the browser and server. These filters include support for inline unit conversion, arbitrary arithmetic, and expose the underlying JavaScript math functions which include, for example, the standard trigonometry functions, logarithms, exponentials, rounding functions, etc. In the Table these custom filters can be used to constrain the set of exoplanets shown and to construct new custom planet properties that can in turn be added as table columns and used in subsequent filters. In the Plotter these custom filters can be used to rapidly construct plots featuring various data cuts.

The Plotter uses the relatively modern HTML canvas tag to implement a fluid, interactive, in-browser plotting environment. We use multiple canvas buffers to make panning and zooming the plot as smooth as possible, even when several complex plots are overlaid on the same figure. The Plotter supports customizable scatter plots and histograms --- scatter plots in particular can display up to four variables simultaneously: the x and y coordinates can each be bound to different quantities as can the marker colors and scales. Of course, the language parser used to construct arbitrary cuts can also be used to specify arbitrary quantities to plot and changes to the plot appear in real-time as they are made. All of this plotting functionality is implemented in JavaScript.

The HTML{\tt  canvas} tag allows us to export the resulting plot directly into the common PNG raster format. To support publication quality output we also allow for vector export in the PDF and SVG formats. To make this possible we implement a secondary SVG plotting backend on the client using the open source Raphael JavaScript library\footnote{{\tt http://raphaeljs.com}}. When the user chooses to export to a vector format the Plotter generates a vector copy of the plot off-screen --- tweaked to look identical to the raster canvas version visible onscreen --- that is then exported to the server where it can be converted to a PDF and sent back to the browser. 

Finally, users can save their plots and tables for later reuse; these saved plots will automatically update to reflect the latest version of the EOD when the user returns to exoplanets.org. This is accomplished without storing any information on the server by, instead, storing the plots/tables in cookies on the user's browser. The benefit here is that we do not need to provide our users with accounts to store any data on our server. The downside is that stored plots and tables will only be available in the same browser that the user created them on and will be lost should the user clear his or her cookies.

\section{Example Plots and the RV-Discovery Sample} 

One of the most useful added values of the EOD is its distinction between planets discovered through radial velocity and those discovered through transit.  This allows for the worst selection effects inherent in both methods to be separated.  We illustrate some of the plotting capabilities of the Exoplanet Data Explorer below with examples of interesting features in the semi-major axis distribution among the RV-discovered planets.  Many of these features have been explored in the literature, especially in \citet{Wright09} and \citet{Wright09d}.

\begin{figure}
\plotone{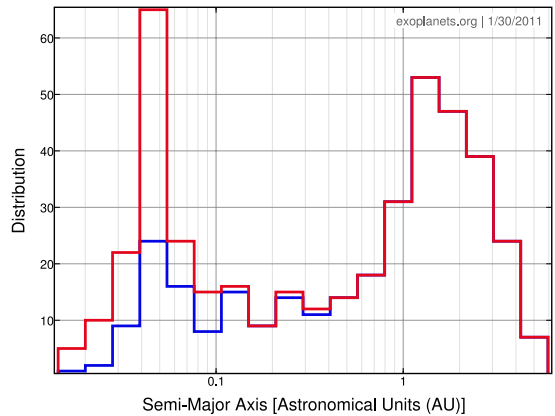}
\caption{Semimajor axis distribution of all planets in the EOD (red) and all RV-discovered planets (blue).  The latter gives a better sense of the true significance of the ``3-day pileup'' compared to longer orbital periods (i.e. $0.1 < a < 0.5$ AU) because the $a$ dependence of the sensitivity of the RV method is weak ($\sim \sqrt{a}$) while the dependence of the transit method sensitivity is much stronger.  \label{loga}}
\end{figure}

Fig.~\ref{loga} shows that the ``3-day pileup'' of close-in planets is significant in the radial velocity sample, appears overwhelming in the overall sample because of the insensitivity of most transit searches to planets with significantly longer period orbits\citep[e.g.][]{Gaudi05a,Gaudi05b}.

\begin{figure}
\plotone{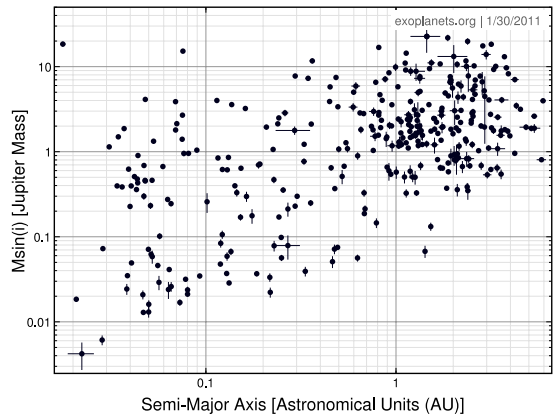}
\caption{$M \sin{i}$ vs. log semimajor axis for all RV-discovered planets.  The lower envelope illustrates the sensitivity of the highest-precision and longest-running surveys. \label{logaMsini}}
\end{figure}

\begin{figure}
\plotone{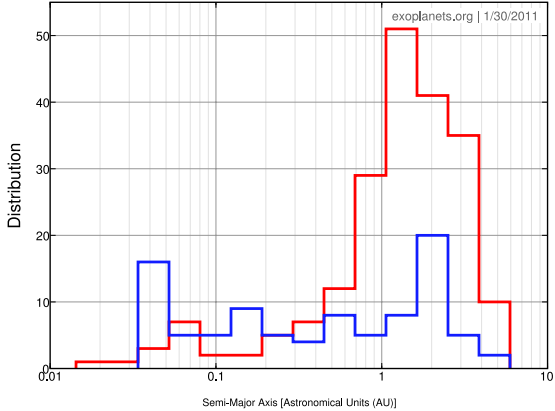}
\caption{Log semimajor axis distribution of RV-discovered super-Jupiters (red) and sub-Jupiters ($0.1 < M \sin{i} < 1 M_{\mbox{\scriptsize Jup}}$,blue).  The ``3-day pileup'' near 0.05 AU does not appear in the super-Jupiter sample.  Note that the sensitivity to sub-Jupiters beyond 0.5 AU falls quickly (see Fig.~\ref{logaMsini}), so the apparently lack of a 1 AU jump in among the sub-Jupiters may be due to lack of sensitivity. \label{logasupersub}}
\end{figure}

Focus on only the RV-discovered planets allows us to explore the nature of the mass-period correlation (Fig.~\ref{logaMsini}).  Comparison of the semi-major axes of super-Jupiters and sub-Jupiters  (Fig.~\ref{logasupersub}) shows that the 3-day pileup is predominantly due to the population of sub-Jupiters, and that super-Jupiters are rarely found in close-in orbits.  The lack of an obvious 1-AU ``jump'' among the sub-Jupiters could easily be due to the difficulty of detecting such planets at such large orbital distances.

\begin{figure}
\plotone{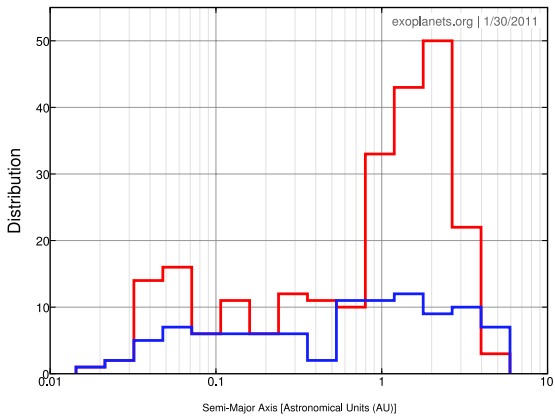}
\caption{Distribution of semimajor axis for all apparently singleton RV-discovered planets (red) and planets in multiplanet systems (blue). These populations follow very different semimajor axis distributions.\label{logamult}}
\end{figure}

Fig.~\ref{logamult} shows that among the multiplanet systems, the semimajor axis distribution is quite distinct:  multiplanet systems are much less likely to include a close-in planet, and there also does not appear to be a 1-AU ``jump'' among the multiplanet systems. 

\begin{figure}
\plotone{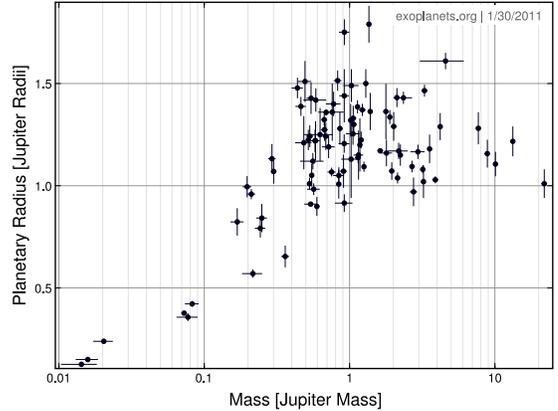}
\caption{Radius vs.\ mass for the known transiting exoplanets.  To illustrate the versatility of the EDE, in this plot the quantity ``Mass'' has been calculated by the web browser as $m \sin{i} / \sin{i}$ from the MSINI and I fields of the EOD, and the uncertainties have been propagated as $\sigma_m = m \sqrt{(\sigma_{\msini} / \msini)^2 + (\sigma_i/\tan{i})^2}$ using the UMISNI and UI fields.  In the browser, each point is clickable and links to that planet's ``one-up'' page.\label{transit}}
\end{figure}

Finally, we illustrate the new transit parameter and uncertainty calcualators.  Fig.~\ref{transit} shows the radius-mass relation for the known transiting systems.  Here, we have calculated the true mass of planets by using the I field of the EOD, and the quantity ``Mass'' is then calculated as {\tt MSINI[mjupiter] / sin(I[rad])}.  We have then chosen to simply propagate the errors in I and MSINI through the error bar calculator as {\tt sqrt((UMSINI[mjupiter]}\verb|^|{\tt2 + (UI[rad] * MSINI[mjupiter] / tan(I[rad]))}\verb|^|{\tt2)) / sin(I[rad])}.  More sophisticated formulae would allow for assymetric errors based on upper and lower limits for I.

\section{Conclusions}

We have made our compilation of robust orbital parameters for all known exoplanets available on {\tt exoplanets.org} through the Exoplanet Orbit Database and the Exoplanet Data Explorer.  The latter is a powerful tool for creating figures and plots for professional and public talks, telescope and funding proposals, for educational purposes in laboratory exercises using authentic data, and for the exoplaration of planet and host star properties generally.  We will continue to update the Database with new planets as they are discovered and the Exolporer with new functionalities.

\acknowledgements
We would like to thank and acknowledge the tireless work of Jean Schneider, whose Extrasolar Planets Encyclopedia is an indispensable reference for all things exoplanetary.  While we have complied every datum in the EOD ourselves from original sources, the Extrasolar Planets Encyclopedia has been a useful check on our numbers and an invaluable clearinghouse of every new planet announcement.

We thank the anonymous referee for a quick and constructive referee's report that improved this manuscript and the Database.  Special thanks go to R. Paul Butler who generated the first peer-reviewed catalog of exoplanets \citep{Butler02}, from which the EOD is descended.    

We thank Scott Gaudi for helpful suggestions that improved this manuscript.  We also thank the many users of the EOD and Data Explorers who sent in edits and suggestions.  We cannot provide a comprehensive list, but such a list would include Michael Perryman, Jean Schneider, Ian Crossfield, Subo Dong, Wes Traub, and Marshall Perrin.  We thank Jason Eastman for a particularly thorough cross-checking of our numbers.

We have made extensive use of the NASA/IPAC/NExScI Star and Exoplanet Database, which is operated by the Jet Propulsion Laboratory, California Institute of Technology, under contract with the National Aeronautics and Space Administration, and NASA's Astrophysics Data System Bibliographic Services.  We thank the NStED administrators, in particular Stephen Kane and David Ciardi, for their assistance and support with the EOD and exoplanets.org, in particular for their help checking database numbers, agreeing to cross-link the websites, and especially for checking, archiving, and providing all published radial velocity data for every exoplanet.

This research has made use of the SIMBAD database, operated at CDS, Strasbourg, France.  This publication makes use of data products from the Two Micron All Sky Survey, which is a joint project of the University of Massachusetts and the Infrared Processing and Analysis Center/California Institute of Technology, funded by the National Aeronautics and Space Administration and the National Science Foundation.  

This work was partially supported by funding from the Center for Exoplanets and Habitable Worlds.  The Center for Exoplanets and Habitable Worlds is supported by the Pennsylvania State University, the Eberly College of Science, and the Pennsylvania Space Grant Consortium.

%\bibliography{references}

\LongTables
\begin{deluxetable*}{lll}
\tablecolumns{3}
\tablecaption{Fields of the Exoplanet Orbit Database}
\tablehead{\colhead{FIELD} & \colhead{Data Type} & \colhead{Meaning}}
\startdata
\label{table}
NAME & String & Name of planet\\ 
STAR & String & Name of host star \\ 
COMP & String & Component name of planet (``b'', ``c'', etc.)\\ 
OTHERNAME & String & Other commonly used star name\\ 
HD & Long Integer & Henry Draper number of star\\ 
HR & Integer & Bright Star Catalog number of star \\ 
HIPP & Long Integer & Hipparcos catalog number of star\\ 
SAO & Long Integer & SAO catalog number of star\\ 
GL & Float & GJ or Gliese catalog number of star\\ 
RA & Double & J2000 Right ascension in decimal hours, Epoch 2000\\ 
DEC & Double & J2000 Declination in decimal degrees, Epoch 2000\\ 
RA\_STRING & String & J2000 Right ascension as a sexigesimal string, Epoch 2000 \\
DEC\_STRING & String & J2000 Declination as a sexigesimal string, Epoch 2000 \\
V & Float & $V$ magnitude\\
BMV & Float & $B-V$ color\\
J & Float & $J$ magnitude\\
H & Float & $H$ magnitude\\
KS & Float & $K_{\mbox{S}}$ magnitude\\
PAR & Float & Parallax in mas\\
UPAR & Float & \\
PER & Double & Orbital period in days\\
UPER  & Float & \\
T0 & Double & Epoch of periastron in HJD\tablenotemark{1}-2440000\\
UT0   & Float & \\
K   & Float & Semiamplitude of stellar reflex motion in m/s\\
UK  & Float & \\
ECC   & Float & Orbital eccentricity\\
UECC   & Float & \\
UECCD  & Float & \\
FREEZE\_ECC & Boolean & Eccentricity frozen in fit?\\
OM   & Float & Argument of periastron in degrees\\
UOM   & Float & \\
TREND & Boolean & Linear trend in fit?\\
DVDT & Float & Magnitude of linear trend in m/s/day\\
UDVDT   & Float & \\
MSINI   & Float & Minimum mass (as calculated from the mass function) in $M_{\mbox{\scriptsize Jup}}$\\
UMSINI    & Float & \\
A    & Float & Orbital semimajor axis in AU\\
UA   & Float & \\
TRANSIT & Boolean & Is the planet known to transit?\\
DEPTH   & Float & $(R_p/R_*)^2$\\
UDEPTH   & Float & \\
UDEPTHD   & Float & \\
T14   & Float & Time of transit from first to fourth contact in days\\
UT14   & Float & \\
TT   & Double & Epoch of transit center in HJD\tablenotemark{1}-2440000\\
UTT  & Float & \\
I  & Float & Orbital inclination in degrees (for transiting systems only)\\
UI  & Float & \\
UID  & Float & \\
R  & Float & Radius of the planet in Jupiter radii\\
UR  & Float & \\
AR  & Float & $(a/R_*)$\\
UAR  & Float & \\
UARD  & Float & \\
B & Float & Impact parameter of transit\\
UB  & Float & \\
UBD  & Float & \\
DENSITY    & Float & Density of planet in g/cc$^3$\\
UDENSITY   & Float & \\
LAMBDA & Float & Projected spin-orbit misalignment\\
ULAMBDA & Float & \\
RMS   & Float & root-mean-square residuals to orbital RV fit\\
CHI2   & Float & $\chi^2_\nu$ to orbital RV fit\\
NOBS   & Integer & Number of observations used in fit\\
NCOMP  & Integer & Number of planetary companions known\\
MULT & Boolean & Multiple planets in system?\\
DISCMETH & String & Method of discovery.  Has value ``RV'' or ``Transit''\\
DATE  & Integer & Year of publication of FIRSTREF\\
MSTAR & Float & Mass of host star\\
UMSTAR & Float & \\
UMSTARD & Float & \\
SPTYPE & String & Spectral type of host star.  Not a fully vetted field.\\ 
BINARY & Boolean & Star known to be binary?\\
FE & Float & Iron abundance (or metallicity) of star.\\
UFE & Float & \\
LOGG   & Float & Spectroscopic $\log g$ (surface gravity) of host star\\ 
ULOGG   & Float & \\ 
TEFF & Float & Effective temperature of host star\\
UTEFF & Float & \\
VSINI & Float & Projected equatorial rotational velocity of star \\
UVSINI & Float & \\
SHK & Float & Mount Wilson Ca {\sc ii} H \& K $S$-value \\
RHK & Float & Chromospheric activity of star as $R^\prime_{\mbox{{\scriptsize HK}}}$\\
JSNAME  & String & Name of host star used in the Extrasolar Planet Encyclopedia \\
ETDNAME & String & Name of hast star used in the Exoplanet Transit Database \\
SIMBADNAME & String & Valid SIMBAD name of host star\\ 
NSTEDID & Long Integer & ID of host star in NStED \\ 
FIRSTREF & String & First peer-reviewed publication of planetary orbit\\ 
FIRSTURL & String & \\ 
ORBREF & String & Peer-reviewed origin or orbital parameters\\ 
ORBURL & String & \\ 
MASSREF & String & Peer-reviewed origin of stellar mass\\ 
MASSURL & String & \\ 
DISTREF & String & Peer-reviewed origin of stellar distance\\ 
DISTURL & String & \\ 
TRANSITREF & String & Peer-reviewed origin of transit parameters\\ 
TRANSITURL & String  \\
BINARYREF& String & Example of peer-reviewed paper mentioning stellar binarity\\ 
BINARYURL & String & 
\enddata
\tablecomments{Fields beginning with U represent uncertainties in the parameter listed before them.  Fields beginning with U and ending with D represent the asymmetric component of these uncertainties, as described in the text.  Fields ending with ``URL'' contain the World Wide Web Uniform Resource Locator to the reference in the corresponding field ending in ``REF''.}
\tablenotetext{1}{The bases for the epoch of transit and periastron passage (JD, HJD, BJD, or others) used in the literature are varied and ocassionally misreported, especially for nontransiting systems.   We have recorded the times given in the original manuscripts, whatever their basis, and plan to report all times consistently in the future.  At present, applications requiring precision to better than several minutes should refer to the TRANSITREF or ORBREF citations.}
\end{deluxetable*}

\end{document}